\documentclass[12pt]{iopart}
\usepackage{graphicx,float}
\usepackage{iopams}
\begin{document}
\title[Dynamical phase transitions in the zero temperature Glauber model]{Rare-event trajectory ensemble approach to study dynamical phase 
transitions in the zero temperature Glauber model}
\author{Pegah Torkaman and Farhad H. Jafarpour}
\eads{\mailto{p.torkaman@basu.ac.ir}, \mailto{farhad@ipm.ir}}
\address{Physics Department, Bu-Ali Sina University, 65174-4161 Hamedan, Iran}
\vspace{10pt}
\begin{abstract}
The dynamics of a one-dimensional stochastic system of classical particles consisting of asymmetric death and branching  
processes is studied. The dynamical activity, defined as the number of configuration changes in a dynamical trajectory, is considered as a 
proper dynamical order parameter. By considering an ensemble of dynamical trajectories and applying the large deviation 
method, we have found that the system might undergo both continuous and discontinuous dynamical phase transitions at critical
values of the counting field. Exact analytical results are obtained for an infinite system. Numerical investigations confirm our analytical 
calculations. 
\end{abstract}
\vspace{2pc}
\maketitle
\section{Introduction}
Phase transitions are remarkable phenomena in both equilibrium and non-equilibrium systems. While traditional thermodynamics can be utilized
to study the static phase transitions and fluctuations associated with configurations of a system, the {\em thermodynamics of trajectories},
sometimes known as Ruelle's thermodynamics~\cite{Ruelle}, can be used to study the dynamical phase behavior. The latter approach 
has recently been adapted to stochastic systems~\cite{LAW}. According to this approach one first considers an ensemble for trajectories of the dynamics over which 
the time-extensive order parameters are defined. The order parameters are physical observables whose fluctuation behavior determines the dynamics 
of the system. In the large deviation limit the probability distributions of the dynamical order parameters are fully captured by large deviation functions 
which play the role of dynamical free-energies, hence they are sometimes called the topological or Ruelle pressure~\cite{Ruelle}.

An important dynamical order parameter, which can be used to classify the various time realizations of the system, is the dynamical activity. 
It is defined as the total number of configuration changes in a trajectory during the observation time interval~\cite{LAW}. This physical observable 
has recently received considerable attention~\cite{HJGC09}. Specifically, it has been employed in studies of the dynamical phase transitions in 
glass former models and lattice proteins~\cite{G09,G1213}.  

In a recent paper~\cite{MTJ14} the authors have studied the fluctuations of a non-entropic current in a one-dimensional 
stochastic system of classical particles which can be considered as a variant of the zero temperature Glauber model. 
In the bulk of the system the particles are subjected to asymmetric branching and death processes. The particles can also 
enter (leave) the system from the left (right) boundary. It is known that, in the long-time limit, the system relaxes into a 
stationary state, where it undergoes a static phase transition from a high-density into a low-density phase, depending on 
the values of the bulk reaction rates. This is characterized as a bulk-induced phase transition. 
In~\cite{MTJ14} the authors have shown that despite the simple nature of the process, the fluctuations of that non-entropic current shows a highly 
non-trivial behavior. While the probability for observing a lower-than-typical current in the steady-state is generated by those 
configurations consisting of a single domain wall, the probability for observing a higher-than-typical current is generated by those 
configurations consisting of multiple domain walls. However, a careful examination of the structure of the large deviation function 
for the probability distribution of this current suggests that the system might undergo dynamical phase transitions. 
This has be inferred from existence of discontinuities in the derivatives of the scaled cumulant generating function of the current which is 
closely related to that of the dynamical activity of the system. 

The main goal of the present paper is to study the dynamical phase transitions in the above mentioned system by investigating the 
behavior of the average activity of the system, as a dynamical observable, below its typical value (i.e. its value in the steady-state). 
We have found that, in the limit of long observation time, the system indeed undergoes both continuous and discontinuous dynamical 
phase transitions. In other words there are four different behaviors for the scaled cumulant generating function of the activity
as a function of the counting field. These different behaviors are associated with the existence of three different dynamical phases in the system. 
In this lower-than-typical activity region, we have characterized different phases according to the configuration of the system at the beginning and the end 
of each trajectory during the observation time. 

At a first-order phase transition point, the system transits  from a phase in which the average activity is solely generated by those 
trajectories whose initial and final configurations are referred to a fully occupied lattice, into another phase where the average activity 
is solely generated by those trajectories whose initial and final configurations are referred to a completely empty lattice and vice versa.
 
At a second-order phase transition point, in contrast, one encounters the following scenario. The average activity in one phase solely 
comes from those trajectories whose initial and final configurations are either referred to a fully occupied or a completely empty lattice while in the 
other phase the initial and the final configurations of the system at the beginning and the end of those trajectories who contribute to the average activity 
are referred to neither a completely empty lattice nor a fully occupied one. 
 
Despite the simplicity of the system studied in this paper, it shows an interesting dynamical phase behavior. While the static phase transition in 
this system is solely determined by the bulk transition rates, the dynamical phase transitions are determined by both the boundary and bulk 
transition rates. 

This paper is organized as follows. In section 2 we start with the mathematical preliminaries. We will then define the model and review the known
results in section 3. The exact expression for the typical activity is given in the section 4. The dynamical phase behavior of the model
is studied in the section 5. The comparison between different dynamical phases is brought in the section 6. In section 7 the large deviation
for the probability distribution of the activity is calculated analytically. The final section is devoted to the concluding remarks.  

\section{Mathematical preliminaries}
In order to present a self-contained paper, we begin with a brief review of the known results on the theory of ensembles of trajectories~\cite{LAW}.
Let us start with a continuous-time Markov process whose configuration space is given by $\{ C \}$. We define $P(C,K,t)$ as the probability of 
being in the configuration $C$ at the time $t$ considering that the system has changed its configuration $K$ times during the time interval $[0,t]$.
The parameter $K$ is in fact the activity of the system at the time $t$. We assume that a spontaneous transition from configuration 
$C$ to $C'$ takes place with a time-independent transition rate $\omega_{C \to C'}$. It is easy to see that $P(C,K,t)$ satisfies the 
following master equation
$$
\frac{d}{dt} P(C,K,t) = \sum_{C'\neq C} \omega_{C' \to C} P(C',K-1,t) - \sum_{C' \neq C} \omega_{C \to C'} P(C,K,t)\; .
$$
Multiplying the above equation by $e^{-sK}$ and summing over all values of  the activity $K \in [0,+\infty)$ we find
$$
\frac{d}{dt} \tilde{P}(C,s,t) = \sum_{C'\neq C} e^{-s}\omega_{C' \to C} \tilde{P}(C',s,t) - \sum_{C' \neq C} \omega_{C \to C'} \tilde{P}(C,s,t)
$$
in which we have defined 
\begin{equation}
\tilde{P}(C,s,t)=\sum_{K=0}^{\infty} e^{-sK} P(C,K,t)\; .
\end{equation}
The parameter $s$ is called the counting field in related literature. Using the quantum Hamiltonian formalism~\cite{Sch} the 
latter master equation can be written as follows
\begin{equation}
\label{SME}
\frac{d}{dt} | \tilde{P}(t) \rangle_s = \tilde{\cal H}_s |\tilde{P}(t)\rangle_s\; .
\end{equation}
Considering the complete basis vector $\{ | C \rangle \}$ the matrix elements of $ \tilde{\cal H}_s $ in this basis are 
$$
\langle C | \tilde{\cal H}_s  | C' \rangle = e^{-s}\omega_{C' \to C} - r(C) \delta_{C,C'}
$$
where $r(C)$, which is the total escape rate from the configuration $C$, is 
$$
r(C)=\sum_{C' \neq C} \omega_{C \to C'} \; .
$$
The formal solution of Eq.~(\ref{SME}) is given by
\begin{equation}
| \tilde{P}(t) \rangle_s = e^{\tilde{\cal H}_s t}  | \tilde{P}(0) \rangle_s \; .
\end{equation}
Since at $t=0$ the activity $K$ is zero, then we have $| \tilde{P}(0) \rangle_s=| P(0) \rangle$ in which
$| P(0) \rangle$ is the probability vector at $t=0$. Note that the probability for being in $C$
at $t=0$ is given by $P(C,K=0,t=0)=\langle C | P(0) \rangle$. It is usually assumed that the system is
in its steady-state at $t=0$; therefore, we choose $| P(0) \rangle=| P^\ast \rangle$ so that
$$
\tilde{\cal H}_{s=0} | P^\ast \rangle=0 \; .
$$
Let us consider an ensemble of trajectories in the configuration space of the system 
during the time interval $[0,t]$. Every member of this ensemble starts, at $t=0$, from a given configuration $C$ with the probability 
$\langle C | P^{\ast} \rangle$ and after the course of time $t$ has elapsed, it might have the activity $K$. We denote the probability 
of having a given activity $K$ at the time $t$ by $P(K,t)$. It is clear that $P(K,t)=\sum_{C}P(C,K,t)$. The moment generating function 
of the activity can now be calculated as follows
\begin{eqnarray*}
\langle e^{-sK} \rangle &=& \sum_{K=0}^{\infty}P(K,t) e^{-sK} \\
&=& \sum_{K=0}^{\infty} \sum_{C} P(C,K,t) e^{-sK} \\
&=& \sum_{C} \tilde{P}(C,s,t) \\
&=& \sum_{C}\langle C | e^{\tilde{\cal H}_s t}  | P^{\ast} \rangle \; .
\end{eqnarray*}
Denoting the right and the left eigenvectors of $\tilde{\cal H}_s$ by $| \Lambda(s) \rangle$ and $\langle \tilde{\Lambda}(s) |$ respectively, we find 
\begin{eqnarray}
\label{ExactGF}
\langle e^{-sK} \rangle &=& \sum_{C}\sum_{\Lambda} \langle C | \Lambda(s) \rangle \langle  \tilde{\Lambda}(s)  | e^{\tilde{\cal H}_s t}  | P^{\ast} \rangle\nonumber \\
& = & \sum_{C}\sum_{\Lambda}  \langle C | \Lambda(s) \rangle \langle  \tilde{\Lambda}(s)  | P^{\ast} \rangle e^{\Lambda(s) t}
\end{eqnarray}
where $\Lambda(s)$'s are the eigenvalue of $\tilde{\cal H}_s$. The above relation can be simplified even further if we assume 
that the large deviation principle holds for a very large observation time. This means that
we asymptotically have
\begin{equation}
\label{Asym}
\langle e^{-sK} \rangle \asymp  \sum_{C} \langle C | \Lambda^{\ast}(s) \rangle \langle  \tilde{\Lambda}^{\ast}(s)  | P^{\ast} \rangle e^{\Lambda^{\ast}(s)t}
\end{equation}
in which $\Lambda^{\ast}(s)$ is the largest eigenvalue of $\tilde{\cal H}_s$. On the 
other hand, $| \Lambda^{\ast}(s) \rangle$ ($\langle  \tilde{\Lambda}^{\ast}(s)  |$) is also the right (left) eigenvector 
of $\tilde{\cal H}_s$ corresponding to the eigenvalue $\Lambda^{\ast}(s)$.  It is known that for the systems 
with unbounded configuration space,~(\ref{Asym}) should be used with care since one of the quantities 
$\sum_{C} \langle C | \Lambda^{\ast}(s) \rangle$ or $\langle  \tilde{\Lambda}^{\ast}(s)  | P^{\ast} \rangle$ might 
diverge~\cite{K98,HRS}. In this case the scaled cumulant generating function of the dynamical observable defined as 
$$
\lim_{t\to \infty} \frac{1}{t} \ln \langle e^{-sK} \rangle
$$ 
is no longer given by the largest eigenvalue of the modified Hamiltonian. This means that the expression~(\ref{ExactGF}) should
be calculated exactly. Note that the infinite dimensionality of the configuration space is only a necessary condition for the violation of~(\ref{Asym}).
In fact it highly depends on the dynamical observable in question. For an explicit example we refer the reader to the models studied in~\cite{HRS}.
In summary, we first calculate the largest eigenvalue $\Lambda^{\ast}(s)$ of the modified Hamiltonian $\tilde{\cal H}_s$ and then check whether any of 
the quantities  $\sum_{C} \langle C | \Lambda^{\ast}(s) \rangle$ or $\langle  \tilde{\Lambda}^{\ast}(s)  | P^{\ast} \rangle$ diverge depending on the value of 
$s$ or the initial distribution defined by $| P^{\ast} \rangle$. The divergence of any of these quantities might result in emergence of new dynamical phases.
Nevertheless, as long as none of the above quantities diverge, the dynamical phase structure of the system is merely determined by the largest eigenvalue 
of the modified Hamiltonian. In this case the role of the initial distribution (which, as we mentioned, is usually assumed to be the steady-state) is marginal.
This means that the probability distribution function of the dynamical observable will be independent of the initial condition. As we will see this is what happens
for the system studied in this paper.

Defining the activity per unit time $k\equiv \frac{K}{t}$ and assuming a large deviation form for $P(k,t)$ 
in the long-time limit , the Legendre transformation of $\Lambda^{\ast}(s)$ gives the large deviation function 
for this probability distribution~\cite{DL98,LS99} 
\begin{equation}
P(k,t) \asymp e^{-t I(k)}
\end{equation}
in which
\begin{equation}
\label{LDF}
I(k)=-\min_s(\Lambda^{\ast}(s)+k s)\; .
\end{equation}
It is assumed that none of the above mentioned quantities in~(\ref{Asym}) diverge; hence, the large deviation function
can be obtained from the Legendre transformation of the largest eigenvalue of the modified Hamiltonian. It is worth mentioning  
that every time the scaled cumulant generating function $\Lambda^{\ast}(s)$ is not differentiable at some points the 
resulting rate function obtained from the Legendre transformation will be linear. One should note that the actual rate function might 
be a nonconvex function instead of a linear function~\cite{T09}. In this case the information embedded in the linear part can only be 
achieved by using other methods such as the direct kinetic Monte Carlo simulation or the population dynamics methods~\cite{GKLT11,TL09}. 

The above mentioned formulation has a simple physical interpretation. Let us consider an ensemble of trajectories in the configuration 
space of the system for which the counting field $s$ has been kept fixed during the course of the time. This ensemble is 
known as $s$-ensemble in related literature. The dynamical partition function of the $s$-ensemble is given by
\begin{equation}
\label{SPF}
Z(s,t)=\langle e^{-sK} \rangle 
\end{equation}
which, as we saw, has a large deviation form in the long-time limit.  From one hand, the largest eigenvalue of 
$\tilde{\cal H}_s$ plays the role of a dynamical free energy whose singularities (or discontinuities of its derivatives with respect to $s$) 
determine the phase behavior of the $s$-ensemble. On the other hand, it can be shown that  the right (left) eigenvector of 
$\tilde{\cal H}_s$ associated with its largest eigenvalue, which has been called $| \Lambda^{\ast}(s) \rangle$ ($\langle \tilde{\Lambda}^{\ast}(s) |$), 
is in fact the probability vector of the final (initial) configuration along a trajectory in the space of configurations, knowing that the value of 
$s$-ensemble average of the activity per unit time has been $\langle k \rangle_s=-\frac{d}{ds}\Lambda^{\ast}(s)$~\cite{Laz13,S09}. 
The $s$-ensemble average of the activity per unit time, or its first cumulant, is defined as
\begin{eqnarray}
\label{SAverage}
\langle k \rangle_s &=& \frac{1}{t} \langle K \rangle_s \nonumber \\
                                  &=& \frac{1}{t}\frac{\langle K e^{-sK}\rangle }{\langle e^{-sK}\rangle}  \\
                                  &=& -\frac{1}{t} \frac{d}{d s} \ln  \langle e^{-sK} \rangle   \nonumber  
\end{eqnarray}
which is equal to $-\frac{d}{ds}\Lambda^{\ast}(s)$ in the large observation time limit.
The $n$'th cumulant of the activity per unit time can be easily obtained by calculating $n$'th derivative of $\Lambda^{\ast}(s)$ with
respect to $s$ times $(-1)^n$. Averages in the $s$-ensemble with $s = 0$ correspond to the steady-state averages or the typical 
values in the steady-state~\cite{G09,FG13}. The typical activity per unit time in the steady-state is given by
\begin{equation*}
\langle k \rangle_{s=0}  =  -\frac{d}{d s} \Lambda^{\ast}(s)\Big |_{s=0}\; .
\end{equation*}
This is the value of the activity per unit time which minimizes $I(k)$ defined in~(\ref{LDF}) and corresponds to the most probable activity
observed in the steady state. 

In summary, the concept of $s$-ensemble enables us to study how the dynamical phase of the system changes
when the activity deviates from its typical value. For a given $s$ the average activity is determined, hence one can 
construct an ensemble of trajectories with that value of the average activity which is not necessarily the typical value in 
the steady state. The necessary information to describe the dynamical phase of the system is embedded in the largest
eigenvalue of the modified Hamiltonian and its corresponding left and right eigenvectors. 

\section{The model: Known results}
The model we are studying in this paper is a one-dimensional stochastic system of classical particles defined on a finite lattice of length $L$
with open boundaries. This is a special case of the model studied in~\cite{J04} which can be considered as a variant of the zero temperature Glauber
model~\cite{SA}. The dynamical rules between two consecutive sites on the 
lattice consist of asymmetric death and branching processes 
\begin{equation}
\label{rules}
\begin{array}{ll}
A \; \emptyset \; \longrightarrow \; \emptyset \; \emptyset \quad \mbox{with the rate} \quad \omega_1 \; ,\\
A \; \emptyset \; \longrightarrow \; A \; A \quad \mbox{with the rate} \quad \omega_2
\end{array}
\end{equation}
in which a particle (vacancy) is denoted by $A$ ($\emptyset$). The particles are also injected into the system from the left 
boundary of the lattice (the first lattice site) with the rate $\alpha$ provided that the target lattice site is already empty. 
They are also extracted from the right boundary of the system (the last lattice site) with the rate $\beta$ provided that it is already occupied. 
The dynamics of the system is irreducible and it has a unique (equilibrium) steady-state. 

It is known that by fine-tuning the microscopic reaction rates, the system might undergo a static bulk-driven phase transition from a high-density 
(for $\omega_1<\omega_2$) to a low-density phase (for $\omega_1>\omega_2$) in the long-time limit~\cite{J04}. Being 
in the steady-state with  $\omega_1<\omega_2$ the lattice becomes almost full of particles except near the right boundary 
where the particles have a chance to leave the system. As the particles leave the system from the right boundary, the second reaction 
in~(\ref{rules}) creates new particles near the right boundary and it maintains the system almost full of particles. One should note that 
the reactions in~(\ref{rules}) do not change the density of particles in a fully occupied region. In summary, the average particle 
density throughout the lattice is equal to $1$ while it falls off exponentially near the right boundary. 
Since the average density of particles in the middle of the lattice is $1$ it is called the high-density phase~\cite{J04}. 
In contrast, being in the steady-state with $\omega_1>\omega_2$ the system is almost empty except near the left boundary where the particles 
still have a chance to enter the system. However, as soon as the particles enter the system the first reaction in~(\ref{rules}) removes them. This
results in an almost empty lattice except near the left boundary where the particle density increased exponentially. One should also note that 
the reactions in~(\ref{rules}) do not change the density of the particles in a completely empty region. Since the density of the particles is
equal to $0$ in the middle of the lattice, this is called a low-density phase~\cite{J04}.

Considering the following basis kets
$$
\vert  \emptyset \rangle = \left( \begin{array}{c}
1\\
0\\
\end{array} \right)\; ,\;\;  
\vert A \rangle=\left( \begin{array}{c}
0\\
1\\
\end{array} \right)\;
$$
which will be used throughout this paper, let us define a product shock measure with a shock front at the lattice site $i$ as
\begin{equation}
\label{SM}
| \{1\}_{i}\{0\}_{L-i} \rangle \equiv 
|A \rangle^{\otimes i}\otimes
|\emptyset \rangle^{\otimes(L-i)} 
\end{equation}
for $0 \le i \le L$. Note that everywhere throughout the paper we will define $\vert \{X\}_{0}\{Y\}_{L} \rangle \equiv \vert \{Y\}_{L} \rangle$. 
The measure $| \{1\}_{i}\{0\}_{L-i} \rangle$ has a simple interpretation. It corresponds to a  
configuration of the system in which the lattice is occupied by particles from the first lattice site up to the $i$'th lattice site.
The rest of the lattice sites are empty.
It is known that in the steady-state the probability vector $| P^{\ast} \rangle$ can be written as a linear superposition of the
product shock measures defined in~(\ref{SM}) with the property that the shock position $i$ performs a biased random walk on the lattice. 
On the other hand, it is known that $| P^{\ast} \rangle$ can also be calculated using a matrix method~\cite{BE07}. 
By assigning the operators $E$ and $D$ to the presence of a vacancy and a particle presented at a given 
lattice site, the steady-state probability for being in a given configuration $\{ \tau \} = \{ \tau_1,\cdots,\tau_L \}$ is given by
\begin{equation}
\label{Weight}
P(\{ \tau \}) \propto \langle\langle W \vert \prod_{i=1}^{L}(\tau_{i} D + (1-\tau_{i}) E) \vert V \rangle\rangle
\end{equation}
in which $\tau_{i}=0$ or $1$ if the $i$'th lattice site is empty or it is already occupied by a particle.
It has been shown that the auxiliary vectors $\vert V \rangle\rangle$ and $\langle \langle W \vert$ besides the operators $E$ and $D$
have a two-dimensional matrix representation given by~\cite{J04}
\begin{equation}
\begin{array}{ll}
\label{Representation}
D=\left( \begin{array}{cc}
0 & 0\\
d & \frac{\omega_{2}}{\omega_{1}}\\
\end{array} \right),\;\;
E=\left( \begin{array}{cc}
1 & 0\\
-d & 0\\
\end{array} \right),\\ \\
\vert V \rangle\rangle=\left( \begin{array}{cc} \frac{-\beta \omega_{2}}{(\omega_{2}-\omega_{1}+ \beta) d\omega_{1}}\\ 
1 \end{array} \right),\;\;
\langle\langle W \vert=\left( \begin{array}{cc} \frac{(\omega_{1}-\omega_{2}+ \alpha)d}{\alpha} & 1 \end{array} \right)
\end{array}
\end{equation}
in which $d$ is a free parameter. This matrix representation allows us to calculate the typical value of any physical quantity 
in the long-time limit such as the typical value of the activity of the system in the steady-state which will be calculated in the next section.

\section{Typical activity in the steady-state}
As we mentioned in the introduction, we consider the dynamical activity, defined as the number of configuration changes in a dynamical trajectory,
as the dynamical order parameter to study the dynamical phase transitions in the system defined by~(\ref{rules}). The reader should note that 
although the reaction rules defined in~(\ref{rules}), besides the injection and extraction rates at the boundaries, are microscopically irreversible processes,
the average entropy production of this system in the steady-state in zero.
It is known that the calculation of the entropy production for a system with microscopically irreversible transitions results in producing of an infinite 
amount of entropy in the environment. During the last couple of years there have been some efforts by different authors to overcome this 
ambiguity~\cite{PJ}; nevertheless, the question on how the entropy production should be defined for these systems is still an open question. 

In what follows we will show that the typical value of the activity in the steady-state $\langle k \rangle_{s=0}$, for which the large deviation 
function $I(k)$ defined in~(\ref{LDF}) is minimum, can be calculated exactly using the matrix method explained in the previous section. The 
typical value of the activity in the steady-state is given by
\begin{equation}
\langle k \rangle_{s=0} =\sum_{C}\sum_{C' \ne C}\omega_{C \to C'} P(C)=\sum_{C} r(C)P(C) \; .
\end{equation}
This can be rewritten as 
\begin{equation}
\begin{array}{ll}
\langle k \rangle_{s=0} = & \alpha \sum_{\{ \tau \}} P(\tau_1=0,\tau_2,\cdots,\tau_L)+
\beta \sum_{\{ \tau \}} P(\tau_1,\tau_2,\cdots,\tau_L=1)  \\ \\
& +(\omega_1+\omega_2) \sum_{\{ \tau \} } \sum_{i=1}^{L-1} P(\tau_1,\cdots,\tau_i=1,\tau_{i+1}=0,\cdots,\tau_L) \; . 
\end{array}
\end{equation}
It turns out that this expression can be calculated exactly using~(\ref{Weight}) and~(\ref{Representation}). After some straightforward  
calculations we find
\begin{equation}
\label{TActivity}
\langle k \rangle_{s=0} =\frac{ (\frac{\omega_2}{\omega_1})^L-1}{
(\frac{\beta -\omega_1+\omega_2}{2\beta\omega_2}) (\frac{\omega_2}{\omega_1})^L-
(\frac{\alpha -\omega_2+\omega_1}{2\alpha\omega_1})}
\end{equation}
In the thermodynamic limit $L \to \infty$ we have
\begin{equation}
\label{average activity}
\langle k \rangle_{s=0} =\left\{
\begin{array}{ll}
\frac{2\alpha\omega_1}{\alpha-\omega_2+\omega_1} & \quad \mbox{for}\quad \omega_1 > \omega_2 \; , \\   \\
2\omega_{1} = 2\omega_{2} & \quad \mbox{for} \quad \omega_{1}=\omega_{2}\; , \\ \\
\frac{2\beta\omega_2}{\beta-\omega_1+\omega_2} & \quad \mbox{for} \quad \omega_1 < \omega_2 \; .
\end{array}
\right.
\end{equation}
Because of the symmetry of the model we will only study the low-density phase where $\omega_1 > \omega_2$. It is easy to 
find the corresponding results in the high-density phase by considering the following transformation
\begin{eqnarray*}
&& \omega_1  \to \omega_2 \\
&& \alpha  \to  \beta \\
&& \mbox{Lattice site number} \;\; i  \to  L-i+1 \; .
\end{eqnarray*}
From now on, we will also drop the $s$-ensemble average subscript and consider $k$ (instead of $\langle k \rangle_s$) as 
the $s$-ensemble average of the activity. The most probable activity or its typical value will also be denoted by $k^{\ast}$.

\section{Dynamical phase behavior of the model}
It is known that the fluctuations of a physical observable in a dynamical system can be captured from analytic properties of the 
largest eigenvalue of its associated modified stochastic Hamiltonian $\tilde{\cal H}_s$ which has been denoted by $\Lambda^{\ast}(s)$. 
Discontinuities in the first and the second derivatives of $\Lambda^{\ast}(s)$ are associated with the first- and the second-order dynamical 
phase transitions. These derivatives are associated with the average and variance of the activity per unit time respectively. 
On the other hand, the positive or negative values of the counting field $s$ favor histories with atypical values of the 
activity~\cite{LAW,G09,FG13,S09,L13}. 

Throughout the forthcoming sections we will mainly concentrate on the positive values 
of the counting filed $s$. This means that we will deal with the $s$-ensemble average of the activity smaller than its typical value. 
As we will see, depending on the values of the microscopic transition rates, the model might undergo both continuous and discontinuous 
dynamical phase transition in this region. 

\subsection{Eigenvalues and Eigenvectors of the modified stochastic Hamiltonian}
Let us start with calculating the eigenvalues and eigenvectors of the modified stochastic Hamiltonian $\tilde{\cal H}_s$ as a $2^L\times 2^L$
irreducible matrix. We will then select the largest eigenvalue and its corresponding eigenvector in the $s \ge 0$ region. 
We have found that the largest eigenvalue of $\tilde{\cal H}_s$ in the $s \ge 0$ region can be obtained by considering the fact that the 
model has a $(L+1)$-dimensional subspace of the configuration space, spanned by the vectors of type~(\ref{SM}) , which 
is invariant under the evolution generated by $\tilde{\cal H}_s$. In other words, acting $\tilde{\cal H}_s$ on any of these 
vectors results in a linear combination of the vectors in the same subspace. More precisely, we have
\begin{equation}
\label{EVQ}
\begin{array}{lll}
\tilde{\cal H}_s\vert \{0\}_{L} \rangle & = & \alpha e^{-s}\vert \{1\}_{1}\{0\}_{L-1} \rangle - \alpha \vert \{0\}_{L} \rangle \; ,  \\
\tilde{\cal H}_s\vert \{1\}_{L} \rangle & = & \beta e^{-s}\vert \{1\}_{L-1}\{0\}_{1} \rangle-\beta \vert \{1\}_{L} \rangle \; ,\\
\tilde{\cal H}_s\vert \{1\}_{i}\{0\}_{L-i} \rangle & = & \omega_{1} e^{-s} \vert \{1\}_{i-1}\{0\}_{L-i+1}  \rangle  
\\ & & + \omega_{2} e^{-s} \vert \{1\}_{i+1}\{0\}_{L-i-1}  \rangle 
\\ & & -(\omega_{1}+\omega_{2})\vert \{1\}_{i}\{0\}_{L-i}  \rangle
\end{array}
\end{equation}
with $1 \le i \le L-1$. One can now easily construct the right eigenvector of $\tilde{\cal H}_s$ by writing 
\begin{equation}
\label{EV}
| \Lambda (s) \rangle =\sum_{i=0}^{L} C_{i}(s) | \{1\}_{i}\{0\}_{L-i} \rangle
\end{equation}
for which 
\begin{equation}
\label{EVE}
\tilde{\cal H}_s | \Lambda (s) \rangle =\Lambda(s) | \Lambda (s) \rangle \; .
\end{equation}
Although this will only give us $L+1$ eigenvalues of $\tilde{\cal H}_s$ out of $2^L$, our careful numerical investigations 
have confirmed that the largest eigenvalue of the $\tilde{\cal H}_s$ in the $s \ge 0$ region lies among these eigenvalues. 
The reader should note that the largest eigenvalue of $\tilde{\cal H}_s$ is equal to zero at $s=0$.

Substituting~(\ref{EV}) in~(\ref{EVE}) and using~(\ref{EVQ}) one finds the equations governing $C_{i}(s)$'s. 
The equations governing $C_{i}(s)$'s consist of a homogeneous linear recursion of order $2$ besides
four boundary recursions. The standard approach to solve such recursions, which is sometimes called a plane 
wave ansatz method, is to consider a solution of the form~\cite{Sch} 
$$
C_{i}(s) = A(z_{1}) z_{1}^{i}+B(z_{2}) z_{2}^{-i} \;\; \mbox{for} \;\; 0\le i \le L \; .
$$
The coefficients $C_{0}(s)$ and $C_{L}(s)$ are also assumed to have the same structure except a multiplication factor.
The relation between the complex parameters $z_1$  and $z_2$ can be obtained from the bulk recursion. The coefficients 
$A$ and $B$ can also be calculated form the boundary recursions. 
This method enables us to calculate both the eigenvalues and eigenvectors of the modified Hamiltonian in the $s \ge 0$ region. 
By substituting these solutions into the recursions and after some straightforward calculations we obtain 
\begin{equation}
\label{Cs}
C_{i}(s)=\eta^{i} \frac{a(z) z^{i}+a(z^{-1})z^{-i}}{(1-\zeta)^{\delta_{i,0}}(1-\xi)^{\delta_{i,L}}} 
\end{equation}
in which we have defined
$$
\eta \equiv \sqrt{\frac{\omega_{2}}{\omega_{1}}} \; ,
\quad \zeta \equiv 1-\frac{\alpha}{\omega_{2}} \; ,
\quad \xi \equiv 1-\frac{\beta}{\omega_{1}}  \; .
$$
From the boundaries recursions one finds
\begin{equation}
\label{As}
\frac{a(z)}{a(z^{-1})}=-\frac{{\cal F}(z,\eta,\zeta)}{{\cal F}(z^{-1},\eta,\zeta)} =-z^{-2L}\frac{{\cal F}(z^{-1},\eta^{-1},\xi)}{{\cal F}(z,\eta^{-1},\xi)}  
\end{equation}
where
\begin{equation}
\label{F}
{\cal F}(x,y,z) \equiv (x+\frac{z}{x})e^{-s} -(yz+\frac{1}{y})\; .
\end{equation}
The eigenvalues are also given by
\begin{equation}
\label{eigenvalues}
\Lambda(s)=-(\omega_{1}+\omega_{2})+e^{-s}\sqrt{\omega_{1} \omega_{2}}(z+z^{-1})
\end{equation}
where the equation governing $z$ is 
\begin{equation}
\label{eq for zs}
z^{2L}=\frac{{\cal F}(z^{-1},\eta,\zeta){\cal F}(z^{-1},\eta^{-1},\xi)}{{\cal F}(z,\eta,\zeta){\cal F}(z,\eta^{-1},\xi)} \; .
\end{equation}
This equation has obviously $2L+4$ solutions. It is clear that if $z$ is a solution, then $z^{-1}$ is also a solution. On the other hand, 
the trivial solutions i.e. $z=\pm 1$ have to be eliminated since they result in two zero eigenvectors. This equation finally gives us $L+1$ 
solutions which correspond to the same number of eigenvalues out of the total $2^L$ eigenvalues of the modified Hamiltonian $\tilde{\cal H}_s$.
As we have mentioned above, it turns out that the largest eigenvalue of $\tilde{\cal H}_s$ in $s\ge 0$ region lies among these $L+1$ 
eigenvalues.

\begin{figure}[t]
\begin{centering}
\includegraphics[width=3in]{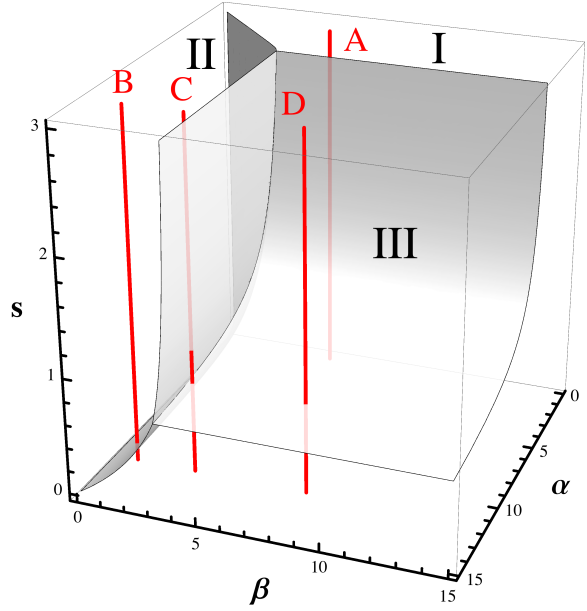}
\caption{\label{Fig1} The dynamical phase diagram of the model in the low-density phase $\omega_1>\omega_2$ for 
$\omega_1=3$ and $\omega_2=1$. Three different phases are denoted by $I$, $II$ and $III$. }
\end{centering}
\end{figure}

For a finite system one can solve~(\ref{eq for zs}) numerically. Nevertheless, one can solve~(\ref{eq for zs}) analytically in the 
thermodynamic limit $L \to \infty$. Assuming $|z|>1$, we have found that in the thermodynamics limit the Eq.~(\ref{eq for zs}) 
has two real solutions and a phase solution. The real solutions are given by
\begin{eqnarray}
\label{Zs1}
z_{1} &=&  {\cal G}(\eta^{-1},\zeta),\; \\
\label{Zs2}
z_{2} &=& {\cal G}(\eta,\xi)
\end{eqnarray}
in which we have defined
$$
{\cal G} (x,y) \equiv e^s (\frac{x}{2}+\frac{y}{2x})+\sqrt{e^{2 s}(\frac{x}{2}+\frac{y}{2x})^2-y} \; .
$$
By substituting these real solutions in~(\ref{eigenvalues}) one finds two eigenvalues $\Lambda_{1}(s)$ and $\Lambda_{2}(s)$ which are 
given by the following expressions
\begin{eqnarray}
\label{eigen1}
\Lambda_{1}(s) &=& \omega_{2} {\cal R}(\eta^{-1},\zeta)\; , \\
\label{eigen2}
\Lambda_{2}(s) &=& \omega_{1} {\cal R}(\eta,\xi) 
\end{eqnarray}
where we have defined 
$$
 {\cal R}(x,y)  \equiv -\frac{(1-y)}{2 y} \left(\sqrt{\left(x^2+y\right)^2-4 e^{-2 s} x^2 y}-x^2+y\right)  \; .
$$
The phase solution of the Eq.~(\ref{eq for zs}) results in an eigenvalue whose maximum value is
given by
\begin{equation}
\label{eigenph}
\Lambda_{\mbox{Ph}}(s)=-(\omega_{1}+\omega_{2})+2\sqrt{\omega_{1} \omega_{2}} e^{-s} \; .
\end{equation}
It turns out that one can also calculate the exact analytical expressions for the normalized right eigenvectors 
of $\tilde{\cal H}_s$ corresponding to the eigenvalues $\Lambda_{1}(s) $, $\Lambda_{2}(s) $ and $\Lambda_{\mbox{Ph}}(s)$
in the large-$L$ limit. The physical interpretations of these vectors will be given in the forthcoming sections. 
By considering~(\ref{EV}) the coefficients for $\vert \Lambda_{1}(s) \rangle$ are
$$
C_i^1(s)=
\frac{ (z-\eta ) (1-\eta  z)  
\left(\eta  \left(1+\zeta z^2\right)
-\eta z^{2 i} \left(\zeta +z^2\right)
-e^s z \left(1+\zeta  \eta ^2\right) \left(1-z^{2 i}\right)\right)}
 { \eta ^{1-i} z^{i+1}
 \left(1-e^s\right) \left(1-z^2\right) \left(1+\zeta  \eta ^2\right)
 (1-\zeta )^{\delta _{i,0}} (1-\xi)^{\delta _{i,L}}}
$$
for $0 \le i \le L$ and $z=z_1$. The coefficients for $\vert \Lambda_{2}(s) \rangle$, which will be called $C_{i}^{2}(s)$, can be obtained 
from the above expression by applying the following transformation
\begin{eqnarray*}
 && \eta \to \eta^{-1} \\
 && \zeta \to  \xi \\
 && \xi \to  \zeta \\
 && i  \to  L-i
\end{eqnarray*}
which also transforms $z_1$ to $z_2$. Finally for  $\vert \Lambda_{\mbox{Ph}}(s) \rangle$ we have 
$$
C_i^{\mbox{Ph}}(s)=\frac{(\eta -1)^2  
   \left(\eta (i+1)-\eta \zeta (i-1)-i e^s \left(1+\zeta  \eta^2\right)\right)}
   {\eta ^{1-i}\left(1-e^s\right) \left(1+\zeta  \eta ^2\right)
   (1-\zeta )^{\delta _{i,0}} (1-\xi )^{\delta _{i,L}}}
$$
for $0 \le i \le L$. 
\subsection{The largest eigenvalue in the $s \ge 0$ region}
In the previous section we diagonalized $\tilde{\cal H}_s$ in an invariant sector in which the largest eigenvalue in the $s \ge 0$ region lies. 
In the thermodynamic limit the largest eigenvalue of $\tilde{\cal H}_s$ highly depends on both the microscopic 
transition rates of the model and the counting field $s$. 

\begin{figure*}[tf]
\begin{centering}
\includegraphics[width=6in]{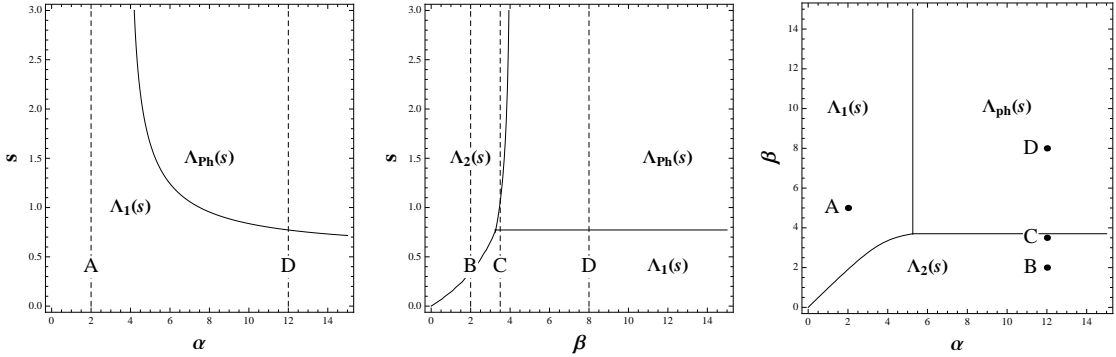}
\caption{\label{Fig2} Three different cross sections of the dynamical phase diagram given in FIG.~\ref{Fig1}.
The leftmost figure is plotted for $\beta=5$ while the middle figure is plotted for $\alpha=10$. Finally, the rightmost
figure is plotted for $s=1.5$. The largest eigenvalue of $\tilde{\cal H}_s$ is clearly specified in each region.}
\end{centering}
\end{figure*}

It turns out that in the low-density phase $\omega_1 > \omega_2$ and for $s \ge 0$, three different regions can be distinguished
depending on the values of $s$, $\alpha$ and $\beta$. In each region the largest eigenvalue of $\tilde{\cal H}_s$ in the 
thermodynamic limit is given by one the expressions $\Lambda_{1}(s)$, $\Lambda_{2}(s)$ or $\Lambda_{\mbox{Ph}}(s)$. The 
boundaries of these three regions are explicitly calculated and given in~\ref{app}. The results are given as a 
$3$-dimensional dynamical phase diagram in FIG.~\ref{Fig1}. In summary, we have
\begin{equation}
\label{LEigenvalue}
\Lambda^{\ast}(s) =\left\{
\begin{array}{ll}
\Lambda_{1}(s) \;\; \mbox{in the region} \;\; I \; ,\\
\Lambda_{2}(s) \;\; \mbox{in the region} \;\; II \; ,\\
\Lambda_{\mbox{Ph}}(s) \;\; \mbox{in the region} \;\; III \; . 
\end{array}
\right.
\end{equation}
The reader can easily check that the first derivative of $\Lambda_{1}(s)$ with respect to $s$ at $s=0$ gives 
the result obtained from the matrix method in~(\ref{average activity}) for $\omega_1 > \omega_2$. For  
$\omega_2 > \omega_1$ one can use the transformation introduced in the previous section. This
can also be obtained from the first derivative of $\Lambda_{2}(s)$ with respect to $s$ at $s=0$.

In order to have a better understanding of the dynamical phase structure of the model, we have specifically 
considered four different points in the space of the parameters defined as follows
 $$
\begin{array}{cccccc}
 & \vline &\omega_1&\omega_2&\alpha&\beta \\  \hline 
A& \vline & 3 & 1 & 2 & 5\\
B& \vline& 3 & 1 & 12 & 2\\
C& \vline & 3& 1 & 12 & 3.5\\
D& \vline & 3 &1 & 12 & 8
\end{array}
$$
These points are plotted as vertical lines parallel to the $s$-axis in FIG.~\ref{Fig1}. In FIG.~\ref{Fig2} we have plotted three different cross sections
of the dynamical phase diagram of the model. These cross sections are three different planes $\beta=5$, $\alpha=10$ and $s=1.5$
in FIG.~\ref{Fig1}.

Along the line $A$ which lies in the region $I$ of FIG.~\ref{Fig1}, we always have $\Lambda^{\ast}(s)=\Lambda_{1}(s)$. 
No dynamical phase transition takes place along this line. Along the line $B$, which is both in the region $I$ and the region $II$, 
a first-order dynamical phase transition takes place at 
\begin{equation}
\label{sc}
s_c=\frac{1}{2} \ln \left(\frac{(\alpha  \omega_{1}-\beta  \omega_{2})^2}{(\alpha -\beta ) \left(-\alpha  \beta  
(\omega_{1}-\omega_{2})+\alpha  \omega_{1}^2-\beta  \omega_{2}^2\right)}\right) \; .
\end{equation}
It turns out that we have $\Lambda^{\ast}(s)=\Lambda_{1}(s)$ for $0<s<s_c$ while $\Lambda^{\ast}(s)=\Lambda_{2}(s)$
for $s > s_c$. The line $C$ goes through the regions $I$, $II$ and $III$. By moving along the line $C$ one might encounter two 
second-order dynamical phase transitions at $s=s_{\alpha}$ and $s=s_{\beta}$ given by the following expressions
\begin{eqnarray}
\label{sasb1}
s_{\alpha} &=& \ln \left( \sqrt{\frac{\omega_{1}}{\omega_{2}}} \frac{\alpha-2\omega_{2}}{\alpha-\omega_{1}-\omega_{2}} \right) \; ,\\
\label{sasb2}
s_{\beta}   &=& \ln \left( \sqrt{\frac{\omega_{2}}{\omega_{1}}} \frac{\beta-2\omega_{1}}{\beta-\omega_{1}-\omega_{2}} \right) \; .
\end{eqnarray}
For $0<s<s_\alpha$ the largest eigenvalue is given by $\Lambda^{\ast}(s)=\Lambda_{1}(s)$, for $s_{\alpha} < s <s_{\beta}$
it is given by $\Lambda^{\ast}(s)=\Lambda_{\mbox{Ph}}(s)$ and for $s > s_{\beta}$ it is given by $\Lambda^{\ast}(s)=\Lambda_{2}(s)$.  
Finally, along the line $D$ a second-order dynamical phase transition takes place at $s=s_{\alpha}$ given by~(\ref{sasb1}).
Along this line and for $0 < s < s_{\alpha}$ we have $\Lambda^{\ast}(s)=\Lambda_{1}(s)$ while for $s>s_{\alpha}$ the largest eigenvalue of 
$\tilde{\cal H}_s$ is given by $\Lambda^{\ast}(s)=\Lambda_{\mbox{Ph}}(s)$. 

As $s \to \infty$, the dynamical phase diagram of the model in terms of $\alpha$ and $\beta$, similar to the one given in the third column of  
FIG.~\ref{Fig2} for a finite $s$, will approach to the following picture. The phase $I$ is limited to the region defined as 
$\alpha < \omega_1+\omega_2$ and $\beta > \alpha$. The phase $II$, on the other hand, is limited to the region 
$\beta < \omega_1+\omega_2$ and $\alpha > \beta$. The phase $III$ is also given by the region 
$\alpha > \omega_1+\omega_2$ and $\beta > \omega_1+\omega_2$. 

\begin{figure}[t]
\begin{minipage}[t]{.5\textwidth}
\includegraphics[width=\textwidth]{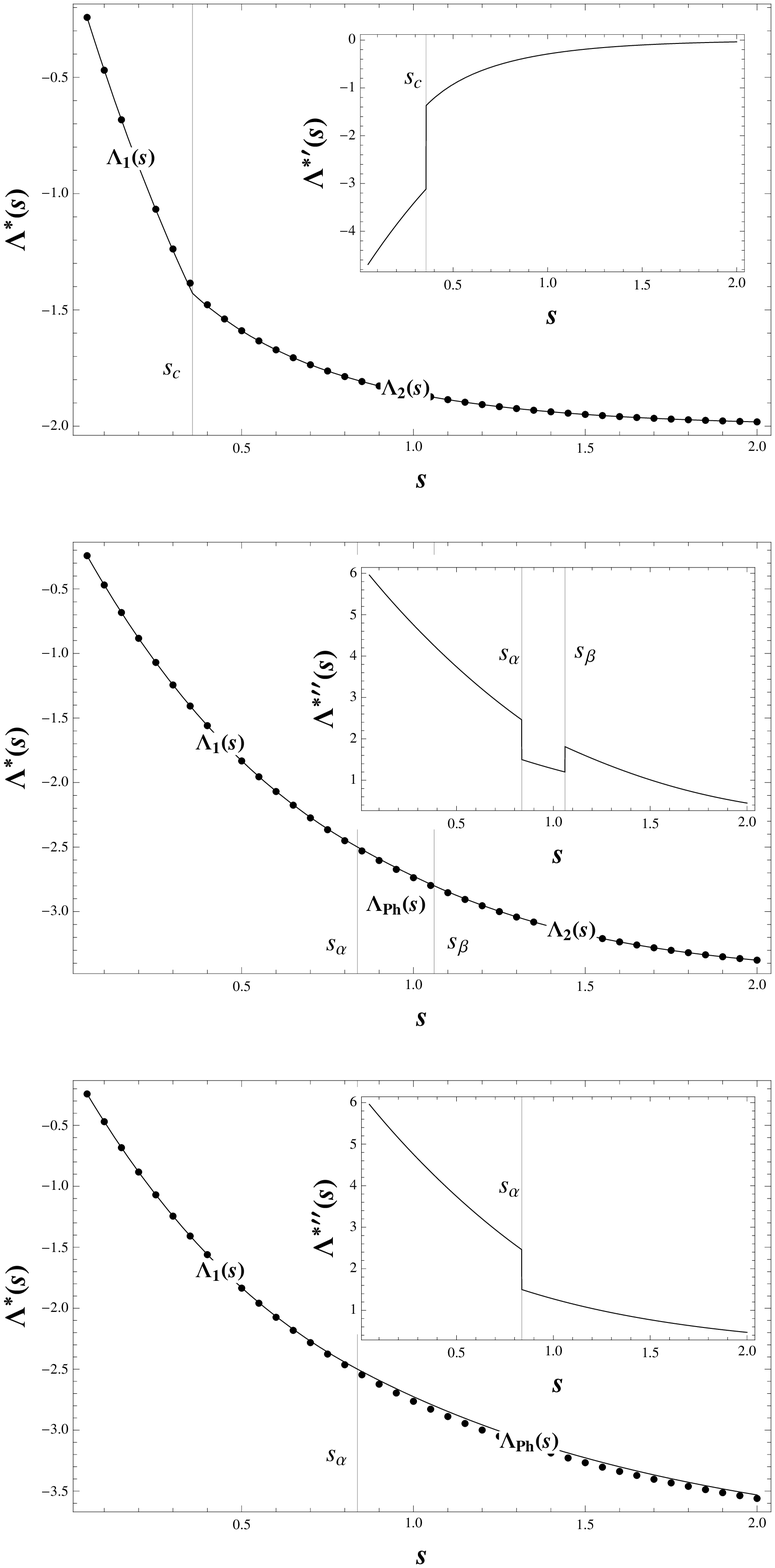}
\caption{The largest eigenvalue of $\tilde{\cal H}_s$ as a function of $s$ for $s \ge 0$ along the three 
lines $B$, $C$ and $D$ from the top to the bottom respectively. The dotted lines are the numerically obtained 
results for a system of length $L=8$ while the full lines are the analytical results obtained in the thermodynamic limit.
The insets show the first and the second derivatives of the largest eigenvalue with respect to $s$ which is obtained 
analytically in the thermodynamic limit.}
\label{Fig3} 
\end{minipage}
\hfill
\begin{minipage}[t]{.5\textwidth}
\includegraphics[width=\textwidth]{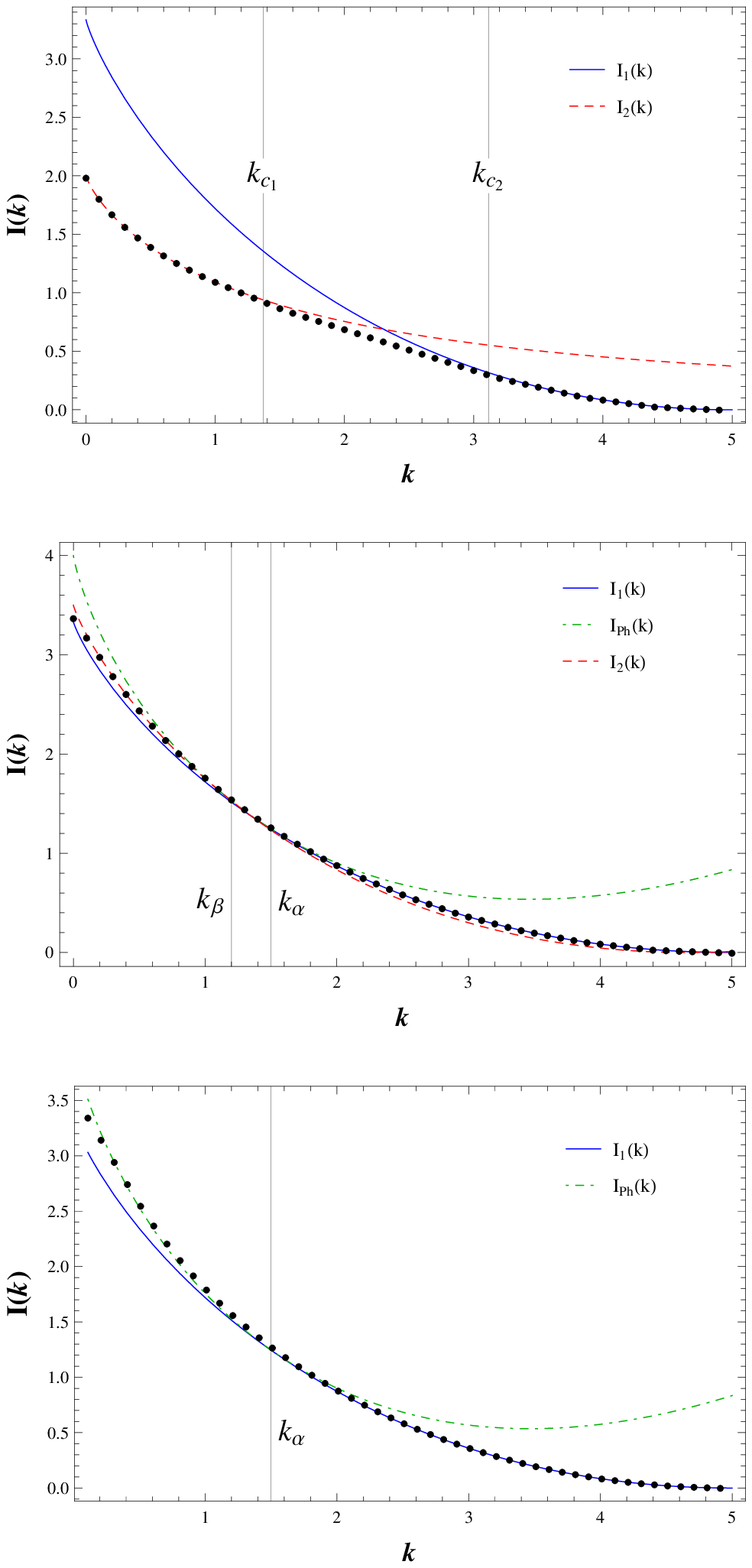}
\caption{The large deviation function $I(k)$ for the probability distribution of the activity along the lines
$B$, $C$ and $D$ from the top to the bottom. The length of the system is $L=8$. The black dotted line is the numerically 
obtained result. The colored lines are the analytical results. For the critical values of the activity $k$ see inside the text.}
\label{Fig4}
\end{minipage}
\end{figure}

The comparison between the numerical and analytical results obtained for the largest eigenvalue of  $\tilde{\cal H}_s$ and its derivatives 
along the three lines $B$, $C$ and $D$ is given in FIG.~\ref{Fig3}. Here we would like to emphasize that by the 
term {\em "numerical calculations"} we mean the exact numerical diagonalization of the modified Hamiltonian and finding its 
largest eigenvalue as a function of $s$ for different system sizes.  As can be seen in FIG.~\ref{Fig3} the numerical and analytical 
results confirm each other. The interpolation function of the numerical solution converges well to the analytical solution. 
The first and the second derivatives of the largest eigenvalue obtained from the analytical calculations are also plotted in the same 
figure. The discontinuities of the first and second derivatives of the largest eigenvalue persist to exist even for small system sizes 
as low as $L=8$ (not shown in the figure). However, they become more prominent (sharper) as the system size increases. The loci of the 
discontinuities of $\Lambda^{\ast}(s)$ obtained from the analytical and numerical approaches are also the same.

\section{Comparison between different dynamical phases}
In this section we would like to comment on the differences between the dynamical phases (or the regions $I$, $II$ and $III$)
introduced in the previous section. Before going into the details let us remind the readers that the steady-state of our model is
unique and that the dynamical phase structure of the model is merely determined by the largest eigenvalue of the modified Hamiltonian. 
Assuming that none of the terms in~(\ref{Asym}) is divergent (which turns to be the case for our model), the large deviation function of the 
system is also unique and independent of the initial configuration of the system and that it can be obtained from the Legendre transformation 
of the largest eigenvalue of the modified Hamiltonian. 

The other point is that, as we have already mentioned, the right and the left eigenvectors of the modified Hamiltonian corresponding to the 
largest eigenvalue give the full information about the possible dynamical trajectories which start from or end to any configuration given that the 
average activity is kept fixed during the observation time. This information is encoded into the modified Hamiltonian itself and is independent 
of the initial configuration of the system provided that none of the terms in~(\ref{Asym}) is divergent. In what follows we will study the general 
properties of the modified Hamiltonian and discuss all possible events regardless of the initial configuration of the system which was assumed 
to be the steady-state.

The final remark is that we are studying the low-density phase $\omega_1 > \omega_2$ under the condition 
$s \ge 0$ which means that our results will be valid for $k <k^{\ast}$. Note also that $s$ and $k$ are related through $k=-\frac{d}{ds}\Lambda^{\ast}(s)$.  
As we mentioned, the right (left) eigenvector of $\tilde{\cal H}_s$ associated with its largest eigenvalue is the probability vector of the 
final (initial) configurations knowing that  $k$ has been kept fixed through the evolution of the system. 

Although the right eigenvector of $\tilde{\cal H}_s$ associated with its largest eigenvalue $| \Lambda^{\ast}(s) \rangle$
can be calculated exactly for any arbitrary positive $s$ in the thermodynamic limit, its left eigenvector $\langle \tilde{\Lambda}^{\ast}(s)|$ 
is more complicated to be calculated. As a matter of fact, as we have already seen, the right eigenvector $| \Lambda^{\ast}(s) \rangle$ 
lies in a small invariant subspace of the total configuration space. In contrast, the left eigenvector $\langle \tilde{\Lambda}^{\ast}(s)|$ should be written 
as a linear combination of $2^L$ properly chosen basis vectors associated with all possible configurations of the system. It turns out 
that finding a closed analytical expression for the coefficients of this expansion for a finite $L$ is a formidable task.

The probability of being in the configuration $\{\tau\}=\{ \tau_1,\ldots,\tau_L \}$ at the end of a trajectory, provided that a fixed $k$ is observed during 
the observation time, can be calculated as follows. Considering the fact that for $s \ge 0$ the final configuration of the 
system can be only one of the states defined in~(\ref{SM}), we define the expression 
\begin{equation}
\label{probfinal}
P_{\mbox{final}}(i|k)\equiv \frac{C_{i}(s)}{\sum_{j=0}^{L}C_{j}(s)} \;\; \mbox{for} \;\; 0 \le i \le L 
\end{equation}
which gives the probability that the final configuration in a trajectory is $\{ \tau_1,\ldots,\tau_L \}=\{ \{1\}_{i}, \{0\}_{L-i} \}$ 
provided that the average activity has been $k$. The coefficients $C_{i}(s)$'s are given in~(\ref{Cs}). On the other hand, the coefficients 
$a(z)$ and $a(z^{-1})$ in~(\ref{Cs}) are given in~(\ref{As}) and the proper $z$ in each region should be obtained from~(\ref{eq for zs}), 
hence for a system of size $L$,~(\ref{probfinal}) can be calculated, in principle,  using~(\ref{Cs})-(\ref{eq for zs}) in each region.

The probability that the initial configuration in a trajectory with a fixed $k$ is $\{\tau\}=\{ \tau_1,\ldots,\tau_L \}$ will be denoted by $P_{\mbox{initial}}(\{\tau\}|k)$.
It turns out that in the thermodynamic limit this quantity can be only calculated analytically for $s\to +\infty$ and $s=0$; however, they can be 
obtained numerically for any $L$ and $s$. 

Let us start with $s=0$ where the only dynamical phase is the region $I$. Regardless of the values of $\alpha$ and $\beta$ the system 
is in the low-density phase and the average activity is given by $k^{\ast}$ in the first line of~(\ref{average activity}). Since at $s=0$ we 
have $|\Lambda^{\ast}(s=0)\rangle=| P^{\ast}\rangle$, using the matrix method one finds that
\begin{equation}
\label{Prob}
P_{\mbox{final}}(i|k^{\ast})=\frac{\left(1-\eta ^2\right) (1-\zeta )^{1-\delta _{0,i}} (1-\xi )^{1-\delta _{L,i}}\eta ^{2 i} }{(1-\xi ) 
\left(1-\frac{\zeta }{\eta ^{-2}}\right)-(1-\zeta ) \left(1-\frac{\xi }{\eta ^2}\right) \eta ^{2 L+2}}
\end{equation}
for $0\le i \le L$. It is clear that in the low-density phase where $\eta <1$ and in the large-$L$ limit, this exponentially decaying function 
is almost zero everywhere except near the left boundary where $i$ is close to zero. This means that the lattice is almost 
deserted or only a few particles are present near the left boundary. The left eigenvector of $\langle \tilde{\Lambda}^{\ast}(s=0)|$ 
can be obtained by noting that $\tilde{\cal H}_{s=0}$ is a stochastic matrix, hence we have 
$$
\langle \tilde{\Lambda}^{\ast}(s=0)|=\Big( 1\;1\;1\; \cdots \; 1\;1\Big)_{1\times 2^{L}}\; .
$$
This implies that the probability for being in any configuration $\{\tau\}$ at the beginning of a trajectory 
is $P_{\mbox{initial}}(\{\tau\}|k^{\ast})=2^{-L}$. This statement simply means that in order to observe
the typical activity during the observation time, the system can start, in principle, from any configuration and there
is no difference from where it starts a trajectory. None of the initial configurations of the system is preferable. There is no 
contradiction here with the fact that in~(\ref{Asym}) the initial configuration of the system can be the steady-state. 

Our numerically exact investigations show that at a finite $s$ the initial configuration of the system at the beginning of a 
trajectory can be basically any arbitrary $\{\tau\}$. However, as $s$ increases toward positive 
infinity the initial configurations of different trajectories can be only one of the states defined in~(\ref{SM}). 

Let us now define $\epsilon \equiv e^{-s}$ where $0 \le \epsilon \le 1$ when $s\ge 0$. Considering the large-$s$ limit corresponds to the very small activities 
and obviously the rare events. Either using the exact results obtained in the previous section in the thermodynamic limit 
or the conventional perturbation method one finds, up to the order $\epsilon^2$
\begin{equation}
\label{EigenS}
\begin{array}{lll}
\Lambda_{1}  (\epsilon) \simeq -\alpha-\frac{\alpha \omega_{1}}{\alpha-\omega_{1}-\omega_{2}}\epsilon^2+{\cal O}(\epsilon^4)\; , \\ \\
\Lambda_{2}  (\epsilon)  \simeq  -\beta-\frac{\beta \omega_{2}}{\beta-\omega_{1}-\omega_{2}}\epsilon^2+{\cal O}(\epsilon^4)\; ,  \\ \\
\Lambda_{\mbox{Ph}} (\epsilon) =  -(\omega_{1}+\omega_{2})+2\sqrt{\omega_{1}\omega_{2}}\epsilon \; .
\end{array}
\end{equation}
As can be seen in~(\ref{EigenS}) the first correction to the largest eigenvalue of $\tilde{\cal H}_\infty$, is 
of order $\epsilon^2$ in the regions $I$ and $II$, while it is of order $\epsilon$ in the region $III$. 
We have also found that, up to the order $\epsilon^2$, the right eigenvector of $\tilde{\cal H}_s$ in the region $I$ is given by
\begin{eqnarray}
\label{lambda1}
|\Lambda^{\ast}(\epsilon)\rangle & \propto& |000\cdots 0 \rangle- \nonumber \\
&& \frac{\alpha \epsilon}{\alpha-\omega_{1}-\omega_{2}} | 100\cdots 0 \rangle+ \nonumber \\
&& \frac{\omega_{1}(\alpha-2\omega_{2})\epsilon^2}{(\alpha-\omega_1-\omega_2)^2} | 000\cdots 0 \rangle+\\
&& \frac{\alpha\omega_{2}\epsilon^2}{(\alpha-\omega_1-\omega_2)^2} | 110\cdots 0 \rangle+ {\cal O} (\epsilon^3) \nonumber
\end{eqnarray}
while its corresponding eigenvalue is given by the first expression in~(\ref{EigenS}). The eigenvector $|\Lambda^{\ast} (\epsilon)\rangle$ 
associated with the largest eigenvalue of $\tilde{\cal H}_s$ in the region $II$, given by the second expression in~(\ref{EigenS}), is 
\begin{eqnarray}
|\Lambda^{\ast} (\epsilon)\rangle & \propto& |1\cdots 111 \rangle- \nonumber \\
&& \frac{\beta \epsilon}{\beta-\omega_{1}-\omega_{2}} | 1\cdots 110 \rangle+ \nonumber\\
&& \frac{\omega_{2}(\beta-2\omega_{1})\epsilon^2}{(\beta-\omega_1-\omega_2)^2} | 111\cdots 1 \rangle+\\
&& \frac{\beta\omega_{1}\epsilon^2}{(\beta-\omega_1-\omega_2)^2} | 1\cdots 100 \rangle + {\cal O} (\epsilon^3) \nonumber \; .
\end{eqnarray}
We have found that in the region $III$, up to the order $\epsilon^0$, the eigenvector of $\tilde{\cal H}_s$ associated with its largest 
eigenvalue, which  is given by $ \Lambda_{\mbox{Ph}} (\epsilon)$ in~(\ref{EigenS}), has the following form
\begin{equation}
\label{pheigenvector}
|  \Lambda^{\ast} (\epsilon)\rangle= \sum_{i=1}^{\infty} i(\eta-1)^2\eta^{i-1} | \{1\}_{i} 0 \cdots 0 \rangle  \; .
\end{equation}
The reader should note that the above expansion does not contain a fully occupied lattice. It is also worth mentioning that for a system 
of size $L$ the largest eigenvalue of $\tilde{\cal H}_{s}$ and its corresponding eigenvector in the region $III$ can be obtained using 
the perturbation method and one finds 
\begin{eqnarray}
\label{III}
\Lambda_{\mbox{Ph}}(\epsilon) &\simeq& -(\omega_1+\omega_2)+2\sqrt{\omega_1\omega_2} \cos (\frac{\pi}{L}) \epsilon +{\cal O} (\epsilon^2)  \; ,\\
|  \Lambda^{\ast} (\epsilon)\rangle &=& \sum_{i=1}^{L-1}\frac{1+\eta^2-2\eta\cos (\frac{\pi}{L})}{1+\eta^L}\times 
\frac{\sin (\frac{\pi i}{L})}{\sin (\frac{\pi}{L})} \times \eta^{i-1}   | \{1\}_{i}\{0\}_{L-i} \rangle
+{\cal O} (\epsilon) \nonumber
\end{eqnarray}
up to the order $\epsilon$ and $\epsilon^0$ respectively. One can readily check that in the thermodynamic limit $L \to \infty$
these expressions converge to the third expression in~(\ref{EigenS}) and~(\ref{pheigenvector}) respectively. 

As $s \to +\infty$, in the region $I$ the largest eigenvalue of $\tilde{\cal H}_\infty$ 
is equal to $-\alpha$. Its corresponding right eigenvector will be denoted by $| 00\cdots 0 \rangle$ which represents an empty lattice. 
It turns out that the left eigenvector of $\tilde{\cal H}_\infty$ with the eigenvalue $-\alpha$ is also given by $\langle 00\cdots 0 |$. This 
means that the trajectories with almost zero activity start from an empty lattice and end to an empty lattice. In the region $II$, the 
right (left) eigenvector of $\tilde{\cal H}_\infty$ is given by $| 11\cdots 1 \rangle$ ($\langle 11\cdots 1 |$) representing a fully occupied lattice. 
The corresponding eigenvalue is $-\beta$. In this phase the zero activity comes from those trajectories which start from a completely 
occupied lattice and also end to the same configuration. The situation in the region $III$ is slightly different. The right eigenvector 
of $\tilde{\cal H}_\infty$ with the eigenvalue $-(\omega_1+\omega_2)$ is highly degenerate although this degeneracy can be easily
removed as we have done in~(\ref{III}) for a finite system. For an infinite system by using~(\ref{pheigenvector}) one finds

\begin{equation}
P_{\mbox{final}}(i|0)=i(\eta-1)^2\eta^{i-1}\;\;\mbox{for}\;\; 1\le i \le \infty \;.
\end{equation}
This distribution is picked around the point $i^{\ast}=| \ln \eta |^{-1}$ and it is almost zero elsewhere. 

In summary, for a finite $s$ and in the large-$L$ limit in the region $I$, it is more probable that the configuration of the system at the beginning 
of a trajectory is an almost empty lattice. Considering the fact that the largest eigenvalue of $\tilde{\cal H}_s$ in this region, given 
by $\Lambda_{1}(s)$ in~(\ref{eigen1}), depends only on $\alpha$ one might conclude that the activity of the system in this 
region is mainly produced by those trajectories during which the system has almost been empty and that only the particle 
injection has generated the activity. In the region $II$ we have found that it is more probable that the initial 
configuration in a trajectory is an almost fully occupied lattice. The largest eigenvalue of $\tilde{\cal H}_s$ 
in this region is given by~(\ref{eigen2}) which clearly depends only on $\beta$. This implies that the right 
boundary or the extraction of particles will play the major role in creating the activity of the system in this region. 
Finally, in the region $III$ it is more probable that the trajectories start from (and end to) those configurations in which the lattice is neither fully 
occupied by the particles nor it is completely empty. The fact that the largest eigenvalue of $\tilde{\cal H}_s$, given by~(\ref{eigenph}), does not depend 
on $\alpha$ and $\beta$ confirms the idea that the activity actually comes from the bulk of the system and the boundaries do not play the role. 

As a closing remark we would like to comment on the dependence of the largest eigenvalue $\Lambda^{\ast}(s)$ in~(\ref{EigenS}) on $\epsilon$.
Let us write the right eigenvector of $\tilde{\cal H}_s$ associated with its largest eigenvalue in each phase as
\begin{equation}
| \Lambda^{\ast}(s) \rangle=| \Lambda^{0} \rangle+\epsilon | \Lambda^{1} \rangle+ \epsilon^2 | \Lambda^{2}\rangle+\cdots
\end{equation}
where $| \Lambda^{0} \rangle =| \Lambda^{\ast}(\infty) \rangle$. The conventional perturbation theory gives the largest eigenvalue as
\begin{equation}
\Lambda^{\ast}(s) = \Lambda^{\ast}(\infty) +\epsilon  \langle \Lambda^{0} | \tilde{\cal H}_s^{\mbox o} | \Lambda^{0} \rangle+ 
\epsilon^2  \langle \Lambda^{0} |\tilde{\cal H}_s^{\mbox o}  | \Lambda^{1} \rangle +\cdots
\end{equation}
in which $\tilde{\cal H}_s^{\mbox o}$ is the off-diagonal part of the matrix $ \tilde{\cal H}_s$.  
In the region $I$ we have $| \Lambda^{0} \rangle=| 00\cdots 0 \rangle$. The first correction to $\Lambda^{\ast}(\infty)=-\alpha$  
is clearly zero since $ \tilde{\cal H}_s^{\mbox o}$ cannot connect an empty lattice to an empty lattice.  However, it does connect
$|10\cdots 0 \rangle$ to $|00\cdots 0 \rangle$ (see~(\ref{lambda1})). This is the reason why the first correction to the eigenvalue in the 
phase $I$ is of the order $\epsilon^2$. The similar reasoning is true for the phase $II$. In the phase $III$, in contrast, the first
correction to the eigenvalue  $\Lambda^{\ast}(\infty)=-(\omega_1+\omega_2)$ is of the order $\epsilon$ since the operator 
$ \tilde{\cal H}_s^{\mbox o}$ can connect $| \Lambda^{0} \rangle$ given in~(\ref{pheigenvector}) to itself.

\section{The large deviation function}
Having the largest eigenvalue of $\tilde{\cal H}_s$ in each region, one can easily use~(\ref{LDF}) to calculate
the {\em convex part} of the large deviation function for the activity of the system along the lines $A$, $B$, $C$ and $D$ analytically.
We have also done the same calculations along the above four lines numerically. As we have already 
mentioned the numerical calculations mean finding the largest eigenvalue of the modified Hamiltonian as a 
function of $s$ numerically and then applying the Legendre transformation to the resulting interpolating function.

Along the line $A$ the largest eigenvalue of $\tilde{\cal H}_s$ is given by $\Lambda_1(s)$ for $0 \le s < +\infty$,
since this line lies completely in the region $I$. There is no discontinuity in $\Lambda_1(s)$ of any type in this 
case. The large deviation function for the probability distribution of activity can be calculated by applying the Legendre 
transformation~(\ref{LDF}) to $\Lambda_1(s)$ given by~(\ref{eigen1}) which results in 
\begin{equation}
I_{1}(k)=\omega_2 {\cal M}(\eta^{-1},\zeta,k\omega_{2}^{-1})
\end{equation}
in which we have defined 
\begin{eqnarray*}
{\cal M} (x,y,z) & \equiv & \frac{1-y}{2y} \Big( y-x^2 +\frac{(y+x^2)^2(1-y)}{zy+\sqrt{z^2y^2+(1-y)^2(y+x^2)^2}} \\
&-& \frac{z}{2} \ln \frac{2x^2( zy+\sqrt{z^2y^2+(1-y)^2(y+x^2)^2})}{z(y+x^2)^2}   \Big). 
\end{eqnarray*}

Along the line $B$ we have a first-order phase transition at $s_{c}$ by the expression in~(\ref{sc}) (see also the first row of FIG.~\ref{Fig3}). 
As it is known~\cite{T09}, the fact that the first derivative of the largest eigenvalue with respect to $s$ is discontinuous at $s_{c}$ will 
result in a linear behavior for the large deviation function. This is actually the nonconvex part of the large deviation function
which cannot be accessed using the Legendre transformation of the largest eigenvalue of the modified Hamiltonian. In summary, the 
large deviation function has three parts: two nonlinear parts $I_1(k)$ and $I_2(k)$ given by
\begin{eqnarray}
I_1(k) &=& \omega_2 {\cal M}(\eta^{-1},\zeta,k\omega_{2}^{-1})\;\; \mbox{for}\;\; k>k_{c_2}\; , \\
I_2(k) &=& \omega_1 {\cal M}(\eta,\xi,k\omega_{1}^{-1})\;\; \mbox{for}\;\; k <k_{c_1}
\end{eqnarray}
in which
\begin{eqnarray}
k_{c_1} &=& \frac{2 \alpha  \omega_{1} (\alpha -\beta ) (-\alpha  \beta  (\omega_{1}-\omega_{2})+\alpha  \omega_{1}^2-\beta 
   \omega_{2}^2)}{| (\alpha +\omega_{1}-\omega_{2}) (\alpha ^2 \omega_{1}^2-\beta ^2 \omega_{2}^2)-2 \alpha  
    \beta  \omega_{1} (\alpha  \omega_{1}-\beta  \omega_{2})| } \; , \\ 
k_{c_2}   &=& \frac{2 \beta  \omega_{2} (\beta -\alpha ) \left(-\alpha  \beta  (\omega_{2}-\omega_{1})-\alpha  \omega_{1}^2+\beta 
   \omega_{2}^2\right)}{| (\beta -\omega_{1}+\omega_{2}) (\beta ^2 \omega_{2}^2-\alpha ^2 \omega_{1}^2)-2 \alpha  
   \beta  \omega_{2} (\beta  \omega_{2}-\alpha  \omega_{1})| }
\end{eqnarray}
and a linear part which connects these two parts. In FIG.~\ref{Fig4} (the first row) we have plotted the large deviation function $I(k)$ as a 
function of $k$ obtained from both analytical and numerical calculations. 
The large deviation function for $k < k_{c_1}$  ($k > k_{c_2}$) is given by $I_{2}(k)$ ($I_{1}(k)$). They both lie on the numerically obtained results. 
For $k_{c_1} < k < k_{c_2}$ it can be seen that the numerical results do not lie on either of the analytical functions. This is the interval where the 
large deviation function is a linear function of $k$.

Along the line $C$ the second derivative of the largest eigenvalue of $\tilde{\cal H}_s$ with respect to $s$ has two discontinuities
at $s_{\alpha}$ and $s_{\beta}$ (see the second row of FIG.~\ref{Fig3}). Since the first derivative of $\Lambda^{\ast}(s)$ is 
continuous along this line, no linear behavior is observed in its corresponding large deviation function. The large deviation
function has three parts along the line $C$. For $0 \le k \le k_{\beta}$ it is given by $I_{2}(k)$ while for $k \ge  k_{\alpha}$
it is given by $I_{1}(k)$ in which 
\begin{eqnarray}
k_{\alpha} &=& \frac{2\omega_2(\alpha-\omega_1-\omega_2)}{\alpha-2\omega_2} \; , \\
k_{\beta}   &=&\frac{2\omega_1(\beta-\omega_1-\omega_2)}{\beta-2\omega_1} \; .
\end{eqnarray}
Finally, for $k_{\beta}\le k \le k_{\alpha}$ the large deviation function is given by 
\begin{equation}
I_{\mbox{Ph}}(k)=\omega_{1}+\omega_{1}-k(1+\ln \frac{2\sqrt{\omega_{1}\omega_{2}}}{k})\; .
\end{equation}
As in the previous case, we have plotted both the analytical and numerical results for the large deviation function along the line $C$ 
in FIG.~\ref{Fig4}. In this case all three parts of the analytical large deviation function lie perfectly on the numerical results. 

As we explained above, along the line $D$ we encounter a second-order phase transition when we move from the region $I$ to 
the region $III$. In terms of the largest eigenvalue of $\tilde{\cal H}_s$ it takes place at $s_{\alpha}$. Using the Legendre 
transformation~(\ref{LDF}) we have found that for $k > k_{\alpha}$ the large deviation function is given by $I_{1}(k)$ while for
$0 \le k \le  k_{\alpha}$ it is given by $I_{\mbox{Ph}}(k)$. As can be seen in FIG.~\ref{Fig4} (the third row) the numerical results 
lie on the theoretical predictions along this line. 

\section{Concluding remarks}

In this paper we have studied the dynamical phase transitions in a one-dimensional stochastic system which can be considered as a
variant of the Zero temperature Glauber model. The bulk reactions
consist of asymmetric death and branching of particles. The particles can enter and leave the system from the boundaries. 
This system is non-ergodic; however, its configuration space is irreducible and that it has a unique (equilibrium) steady-state. In the steady-state
the system undergoes a static phase transition in the thermodynamic limit which is controlled by the bulk reaction rates. Considering the activity 
of the system as a dynamical order parameter, we have found that the system might undergo a dynamical phase transition in the thermodynamic limit which 
is  determined by both the boundary and bulk transition rates. It turns out that the dynamical phase diagram has three different regions or phases. 
The physical properties of each phase is studied in detail in the thermodynamic limit. We have found that the activity of the system is generated either by the 
reactions at the boundaries (in the regions $I$ and $II$) or by the bulk reactions of the system (in the region $III$). 

We have started with a system of length $L$ which has clearly a bounded configuration space. Taking the limit $t \to \infty$ one finds 
that the scaled cumulant generating function is given with the largest eigenvalue of the modified Hamiltonian. We will then take the 
thermodynamic limit $L \to \infty$.  It seems that in the thermodynamic limit the configuration space of our system becomes infinitely 
large; however, we have carefully checked that neither $\sum_{C} \langle C | \Lambda^{\ast}(s) \rangle$ 
nor $\langle  \tilde{\Lambda}^{\ast}(s)  | P^{\ast} \rangle$ diverge (at least for $s \ge 0$). 
From one hand, since $ | \Lambda^{\ast}(s) \rangle$ is calculated 
exactly in all three dynamical phases in the thermodynamic limit, one can easily check that 
$\sum_{C} \langle C | \Lambda^{\ast}(s) \rangle$ never diverges. On the other hand, if we consider   
$\langle  \tilde{\Lambda}^{\ast}(s)  | P^{\ast} \rangle$ as a power series of $\epsilon$ 
we can easily see that  for $\epsilon =0$  the expression $\langle  \tilde{\Lambda}^{\ast}(\infty)  | P^{\ast} \rangle$ is clearly 
finite and equal to the probability of having a completely empty or a fully occupied lattice in the phases $I$ and $II$ respectively. 
Our exact analytical calculations in the phase $III$ show that $\langle  \tilde{\Lambda}^{\ast}(\infty) | P^{\ast} \rangle$ is also convergent.
For $\epsilon=1$ one finds that  $\langle  \tilde{\Lambda}^{\ast}(s)  |=\sum_{C} \langle C | $ which results in
$\langle  \tilde{\Lambda}^{\ast}(s) | P^{\ast} \rangle=1$.

In summary, the expression~(\ref{Asym}) is not divergent regardless of the value of 
$s$ or the initial configuration of the system.  Hence, the dominant eigenvalue of the modified Hamiltonian of the system generates, through 
the G\"artner-Ellis Theorem, a large deviation function for the activity of the system~\cite{T09}. 
We have  calculated the convex part of the large deviation function for the probability distribution function of the activity 
in each phase. Our analytical calculations are compared with the results obtained from numerical diagonalization of the 
modified Hamiltonian.  

In~\cite{HPS} the authors have studied the dynamics of instantaneous condensation in a single-site zero-range process conditioned
on an atypical current. The effective dynamics of their model maps to a biased random walk on a semi-infinite lattice. Since the dynamics
of our model for $s>0$ (the activities lower-than-typical value) can be explained in terms of the dynamics of a single shock moving on a finite
lattice with reflecting boundaries, it can be considered as a conserved zero-range process with two sites and $L$ particles.  
From this point of view, the condensation (accumulation of the particles in a single site) in this zero-range process can be seen in the 
dynamical phases $I$ and $II$ in the limit $s \to \infty$ where the activity goes to zero. Note that this phenomenon is happening in the
low-density phase $\omega_1 > \omega_2$ where the system is typically empty. 

Finally, in~\cite{BS} the authors have studied the finite-time evolution of shocks and antishocks in the asymmetric
simple exclusion process on a ring conditioned on an atypically low particle current. It seems that the lower-than-typical activity 
region in our model ($s>0$) is governed by the evolution of a single shock which performs random walk on the lattice.
The higher-than-typical activities, on the other hand, should be generated by the evolution of multiple shocks in the system.   
One should note that the antishocks in our model do not evolve in time under the dynamical rules~(\ref{rules}) unless one
considers the reactions $\emptyset A \to AA$ and $\emptyset A \to \emptyset \emptyset$. 

There are still many open problems that can be studied separately. Most of our calculations are performed in the thermodynamic limit. 
It would be interesting to study the finite-size effects i.e. the dependence of the eigenvalues and the eigenvectors of the modified Hamiltonian of 
the system on the length of the lattice $L$ in all three dynamical phases~\cite{ABN}. On the other hand, the case $s < 0$ which corresponds to 
the average activity above the typical value has not been studied in this paper and needs a careful and detailed study. The largest eigenvalue 
of the modified Hamiltonian for negative values of $s$ generates the large deviation for the fluctuations of the activity above its typical value. 

\appendix
\section{\label{app} The boundaries of the regions $I$, $II$ and $III$}
As we mentioned, depending on the values of the microscopic transition rates and the counting field the largest eigenvalue of 
 $\tilde{\cal H}_s$ is given by a different expression in the thermodynamic limit. Defining 
$$
U(s,\zeta) \equiv \frac{e^{2 s}\left(1-\eta ^2\right)  \left(\sqrt{\left(\zeta +\eta ^{-2}\right)^2-
4 \zeta  e^{-2 s}\eta ^{-2}}+\zeta \right) -e^{2 s }\left(\eta ^{-2}+1\right)+2}{2 \left(e^{2 s} \left(\zeta(\eta^{-2}-1) -\eta ^{-2}\right)+1\right)} 
$$
these regions are:
\begin{itemize}
\item{Region $I$: In this region $\Lambda^{\ast}(s)=\Lambda_{1}(s)$. This region is defined by 
\begin{eqnarray*}
&& s <\ln \eta^{-1} \; , \frac{1-\xi }{1-\zeta }>U(s,\zeta)  \\
&& s > \ln \eta^{-1} \; , 1-\zeta <\frac{2-\left(\eta +\eta^{-1} \right) e^s}{1-\eta  e^s}\; , \frac{1-\xi }{1-\zeta }>U(s,\zeta) 
\end{eqnarray*}
}
\item{Region $II$: In this region $\Lambda^{\ast}(s)=\Lambda_{2}(s)$. This region is defined by 
\begin{equation*} 
1-\xi < \frac{2-\left(\eta +\eta^{-1} \right) e^s}{1-\eta^{-1}  e^s}\; , \frac{1-\xi }{1-\zeta } <  U(s,\zeta) 
\end{equation*}
}

\item{Region $III$: In this region $\Lambda^{\ast}(s)=\Lambda_{\mbox{Ph}}(s)$. This region is defined by 
$$
s > \ln \eta^{-1}\; , 
1-\zeta  > \frac{2-\left(\eta +\eta^{-1} \right) e^s}{1-\eta  e^s}\; , 
1-\xi  > \frac{2-\left(\eta +\eta^{-1} \right) e^s}{1-\eta^{-1}  e^s} \; .
$$
}
\end{itemize}

 \section*{References}



\begin{thebibliography}{99}
\bibitem{Ruelle} D. Ruelle, {\it Thermodynamic Formalism} (Addison-Wesley, Reading, Massachusetts, 1978);
                 J. P. Eckmann, D. Ruelle, Rev. Mod. Phys. 57, 617 (1985).

\bibitem{LAW}  V. Lecomte, C. Appert-Rolland and F. van Wijland, J. Stat. Phys. 127, 51 (2007).    

\bibitem{HJGC09} 
C. Maes and K. Neto\v{c}n\'{y}, Europhys. Lett. 82, 30003 (2008);
T. Bodineau and R. Lefevere, J. Stat. Phys. 133, 1 (2008);
L. O. Hedges, R. L. Jack, J. P. Garrahan, and D. Chandler, Science 323, 1309 (2009);
M. Baiesi, E. Boksenbojm, C. Maes, B. Wynants, J. Stat. Phys.139, 492 (2010);
E. Pitard, V. Lecomte, and F. Van Wijland, Europhys. Lett. 96, 184207 (2011).

\bibitem{G1213} 
V. Lecomte, J. P. Garrahan and F. van Wijland, J. Phys. A: Math. Theor. 45, 175001 (2012);
J. M. Hickey, C. Flindt and J. P. Garrahan, Phys. Rev. E. 88, 012119 (2013);
A. S. J. S. Mey, P. L. Geissler and J. P. Garrahan, Phys. Rev. E 89, 032109 (2014).           

\bibitem{G09} J. P. Garrahan, R. L. Jack, V. Lecomte, E. Pitard, K. van Duijvendijk and F. van Wijland, J. Phys. A: Math. Theor. 42, 075007 (2009);
              M. Gorissen, J. Hooyberghs, and C. Vanderzande, Phys. Rev. E 79, 020101 (2009).

\bibitem{MTJ14} S. R. Masharian, P. Torkaman and F. H. Jafarpour, Phys. Rev. E 89, 012133 (2014).

\bibitem{Sch} G. M. Sch\"utz, \textit{Phase transitions and critical phenomena} (Academic,London,2001), vol. 19, p.3.

\bibitem{HRS} R. J. Harris, A. R\'{a}kos, and G. M. Sch\"{u}tz, Europhys. Lett. 75, 227 (2006);
           A. R\'{a}kos and R. J. Harris, J. Stat. Mech. P05005 (2008).

\bibitem{K98} Kurchan J., J. Phys. A: Math. Gen. 31, 3719 (1998).

\bibitem{DL98} B.Derrida, J. L. Lebowitz, Phys. Rev. Lett. 80, 209 (1998).  
                               
\bibitem{LS99} J. L. Lebowitz, H. Spohn, J. Stat. Phys. 95, 333 (1999).

\bibitem{T09} H. Touchette, Phys. Rep. 478, 1 (2009).

\bibitem{GKLT11} C. Giardina, J. Kurchan, V. Lecomte, and J. Tailleurm,  J. Stat. Phys. 145, 787 (2011).
\bibitem{TL09} J. Tailleur and V. Lecomte, Proceedings of the 10th Granada Seminar on Computational and Statistical Physics, 
(Granada, 2008), AIP Conf. Proc. 1091 212-219 (2009).

\bibitem{Laz13} A. Lazarescu, Ph.D. thesis, Universit\'e Pierre et Marie Curie - Paris VI, 2013. arXiv:1311.7370

\bibitem{S09} D. Simon, J. Stat. Mech. : Theor. Exp.  P07017 (2009). 

\bibitem{FG13} C. Flindt, J. P. Garrahan, Phys. Rev. Lett. 110, 050601 (2013). 

\bibitem{J04} F. H. Jafarour, Physica A 339, 369 (2004).

\bibitem{SA} G. M. Sch\"utz; Eur. Phys. J. B5  589 (1998),
M. Khorrami and A. Aghamohammadi, Phys. Rev. E 63, 042102 (2001).

\bibitem{BE07} R. A. Blythe, M. R. Evans, J. Phys. A Math. Theor. 40, R333 (2007).

\bibitem{PJ} D. ben-Avraham, S. Dorosz, and M. Pleimling, Phys. Rev. E 84, 011115 (2011), 
 S. Zeraati, F. H. Jafarpour, and H. Hinrichsen, J. Stat. Mech. L12001(2012). 

\bibitem{L13} A. Lazarescu, J. Phys. A: Math. Theor. 46, 145003 (2013).

\bibitem{HPS} R. J. Harris, V. Popkov and G. M. Sch\"utz, Entropy 15(11), 5065 (2013).

\bibitem{BS} V. Belitsky and G. M. Sch\"utz, J. Stat. Phys. 152, 93 (2013).

\bibitem{ABN} 
C. Appert-Rolland, B. Derrida, V. Lecomte and F. van Wijland, Phys. Rev. E 78, 021122 (2008);
T. Bodineau, V. Lecomte and C. Toninelli, J. Stat. Phys. 147, 1 (2012);
T. Nemoto, V. Lecomte, S.-i. Sasa and F. van Wijland, J. Stat. Mech. P10001 (2014).

\end{thebibliography}
\end{document}